\documentclass[aps,showpacs,twocolumn]{revtex4}
\usepackage{graphicx}
\begin{document}
\title{Non-resonant nonlinear coupling of magnetohydrodynamic waves in inhomogeneous media}
\author{V.M. Nakariakov, D. Tsiklauri and T.D. Arber}
\affiliation{Physics Department,
University of Warwick, Coventry, CV4 7AL, England. E-mail:
valery@astro.warwick.ac.uk}
\date{\today}
\begin{abstract}
A new mechanism for the enhanced generation of compressible 
fluctuations by Alfv\'en
waves is presented. A strongly nonlinear regime of  Alfv\'en
wave phase-mixing is numerically simulated in a 
one-dimensionally inhomogeneous plasma 
of  finite temperature.
It is found that the inhomogeneity of the medium determines the efficiency
of nonlinear excitation of magnetoacoustic waves. The level of the compressible 
fluctuations is found to be higher (up to the factor of two) in inhomogeneous regions. 
The amplitude of the generated magnetoacoustic
wave can reach up to 30\% of the source Alfv\'en wave amplitude, and this value is
practically independent of the Alfv\'en wave amplitude and the steepness of Alfv\'en speed profile.
The highest amplitudes of compressible disturbances are reached in plasmas with
$\beta$ of about 0.5. The further growth of the amplitude of compressible 
fluctuations is depressed by saturation.
\end{abstract}

\maketitle

Magnetohydrodynamic (MHD) waves play an important role in the dynamics
of laboratory, space and astrophysical plasmas (e.g.
\cite{mhdrev}). The simultaneous existence of several MHD wave
modes,  three, at least, in the simplest case, stimulates the
interest to processes of  mode coupling and interaction. In
particular, the interaction of linearly incompressible Alfv\'en waves
with compressible magnetoacoustic waves is essential in problems
such as the heating of the open corona of the Sun and acceleration of the
solar wind. Indeed,  Alfv\'en waves are often named as the
carriers of the energy from the lower layers of the solar
atmosphere to the corona. Also, in the inner heliosphere, Alfv\'en
waves represent the main component in MHD turbulence
\cite{tu}. However, compressible waves are subject
to more efficient dissipation, as they decay on volume viscosity,
not shear as to incompressible Alfv\'en waves, and the difference
of these two viscosities can be several orders of
magnitude. Moreover, this mechanism 
allows energy to be transported across field lines in
contrast to Alfv\'en waves which can only transport
energy along the field.
Also, compressible waves perturb the
plasma density and consequently can be detected, again in contrast
to Alfv\'en waves, with imaging telescopes.

A classical example of such interaction is the decay instability
of Alfv\'en waves, connected with the {\it resonant} three-wave
interaction of Alfv\'en and magnetoacoustic waves \cite{3wi}. The
efficiency of this phenomenon is determined by the amplitudes of
the interacting waves. However, this mechanism works only for
quasi-periodic (perhaps wide-spectrum, \cite{malara2000}) waves, and is
not so prominent for wave pulses that could be generated e.g. by
some explosive events such as solar flares, coronal mass ejections, etc. 

In contrast, {\it non-resonant}
mechanisms for compressible fluctuation excitation are independent
of the wave spectrum and can be efficient even for short wave pulses.
An example of {\it non-resonant} MHD wave interaction is the nonlinear
excitation of longitudinal magnetoacoustic perturbations by
nonlinear elliptically polarized Alfv\'en waves through the
ponderomotive force. This phenomenon is actually the generation of
the second harmonic and, in contrast with three wave resonant
interaction, and does not require the daughter wave to be present in
the system from the very beginning. Also, the efficiency of this
process is restricted by the source wave amplitude, but does not
require the source wave to be harmonic. 
Another parameter
affecting the efficiency is $\beta$, the thermal to magnetic
pressure ratio. 
The generation of the compressible perturbations
leads to self-interaction and consequent steepening of the
Alfv\'en wave, whose evolution is described by the Cohen-Kulsrud
equation \cite{cke}.

Yet another non-resonant mechanism for the generation of compressible
fluctuations arises when the Alfv\'en speed varies across the
magnetic field. Initially plane, linearly polarized Alfv\'en waves
become oblique and sharp gradients in the direction across the
field are secularly generated. This phenomenon is 
known as 
 Alfv\'en wave phase mixing, and it has been 
intensively studied in the context of the solar coronal heating
problem \cite{awpm} and often seen in full MHD numerical simulations
(e.g. \cite{malara}, \cite{numpm}). In the compressible regime, phase mixing of
Alfv\'en waves is accompanied by nonlinear generation of fast
magnetoacoustic waves \cite{nrm}.
In this phenomenon, the efficiency of {\it nonlinear coupling} is
dramatically affected by the {\it inhomogeneity} of the medium. To
illustrate this, consider a low-$\beta$ plasma with the straight and
uniform magnetic field ${\bf B}_0 = B_0 {\bf e_z}$, where $B_0$ is
the absolute value of the unperturbed field and ${\bf e_z}$ is the
unit vector in $z$-direction. The plasma mass density $\rho_0(x)$
varies across the field, in the $x$-direction. Thus, the
Alfv\'en speed $C_A(x) = B_0 [4\pi \rho_0(x)]^{-1/2}$ also varies
across the field. In this geometry, Alfv\'en waves (perturbing
$V_y$ and $B_y$) and fast magnetoacoustic waves (perturb $V_x$,
$B_x$, $B_z$ components of the bulk velocity, magnetic field
and the mass density $\rho$) are linearly 
coupled if the Alfv\'en waves have a component of $\vec k$ in the $y$-direction
\cite{ddy}. However, when $\partial/\partial y$ tends to zero, the
waves become {\it linearly} decoupled from each other and  in
this case, the effect of nonlinear coupling becomes pronounced.
The initial stage of  fast magnetoacoustic wave generation by 
 transverse gradients in weakly nonlinear Alfv\'en waves is described by the equation
\begin{equation}
\frac{\partial^2}{\partial t^2}
B_x - C_A^2(x)\left( \frac{\partial^2}{\partial x^2} +
\frac{\partial^2}{\partial z^2} \right)B_x =  \frac{B_0}{8\pi \rho_0(x)} 
\frac{\partial^2}{\partial x
\partial z} B_y^2.
\label{fwe}
\end{equation}
The right handside of Eq.~(\ref{fwe}) in the
Fourier transform domain is proportional
to $k_z k_x A^2$, where $k_z$ and $k_x$ are the components of the wave number in the longitudinal and
transverse directions respectively and $A$ is the Alfv\'en wave amplitude. 
Consequently,
even in the case of homogeneous Alfv\'en speed ($C_A = const$),
a non-plane ($k_x \neq 0$) Alfv\'en wave  generates compressible perturbations,
which, according to the left hand side of Eq.~(\ref{fwe}), propagate isotropically.

In the weakly nonlinear regime the Alfv\'en wave steepening can be neglected and the linear
solution can be used,
\begin{equation}B_y = A f[z - C_A(x) t],
\label{aws}
\end{equation}
 where $f$ is an arbitrary smooth
function prescribed by the initial conditions. 

With an inhomogeneous Alfv\'en speed,
the Alfv\'en wave is subject to phase mixing and $k_x\to\infty$
as time progresses. 
Consequently, the 
phase mixing {\it amplifies} the right hand side term of Eq.~(\ref{fwe}) and affect MHD wave
coupling. Indeed, 
with the Alfv\'en wave given by Eq.~(\ref{aws}),
the right hand side of Eq.~(\ref{fwe}) becomes
\begin{equation}
{\rm RHS}(\ref{fwe}) \propto  A^2 \frac{d C_A(x)}{d\,x} k_z^2 t
\label{secg}
\end{equation} 
and, consequently, it grows {\it secularly} in time, proportionally to the product of
the Alfv\'en wave amplitude and the inverse characteristic spatial
scale of the medium inhomogeneity (see \cite{nrm}, \cite{gert} for details).  
The secularity in the nonlinear generation of the fast waves is entirely connected with
the inhomogeneity of medium. Thus, even if the Alfv\'en wave
amplitude is weak, {\it the generation of fast wave can be
dramatically amplified by the medium inhomogeneity}.
In the homogeneous case, the same ponderomotive force can also
generate perpendicular compressible fluctuations if the Alfv\'en
wave is initially non-plane. However, in this case, the
perpendicular gradients in the Alfv\'en wave remain as prescribed
by the initial conditions and the generation of compressible
fluctuations is limited by the initial shape of the wavefront. The
induced longitudinal motions, also generated by the Alfv\'en wave,
and do not grow secularly either.

\begin{figure}
\includegraphics[width=3.5in]{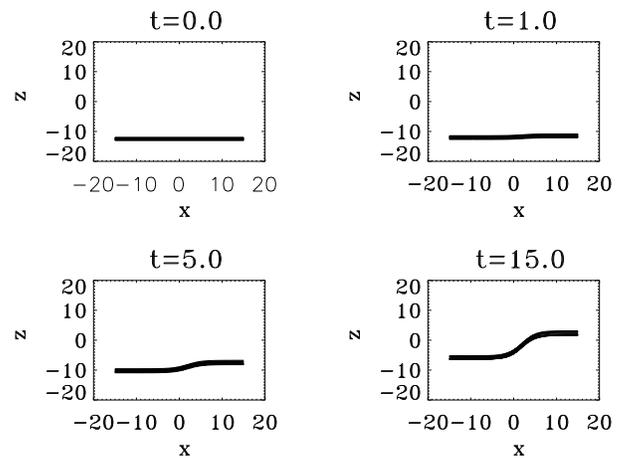}%
\caption{Contourplots of the evolution of
 an initially plane single Alfv\'enic pulse in 
a plasma with an Alfv\'en speed varing in the $x$-direction. 
The unperturbed magnetic field is straight and has the $z$-component only.
Here $A=0.5$, $\lambda=0.31$, $\beta=2$. 
\label{fig1}}
\end{figure}

\begin{figure}
\includegraphics[width=3.5in]{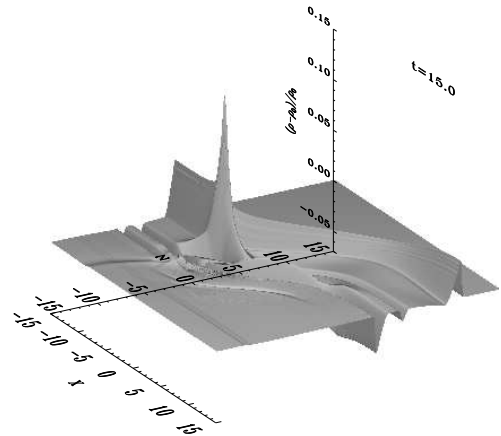}%
\caption{A snapshot of the density perturbation
at $t=15$, generated by an Alfv\'enic pulse of the relative amplitude 0.5,
in a plasma with $\beta = 2$ and inhomogeneity parameter $\lambda=0.31$.
\label{fig2}}
\end{figure}

\begin{figure}
\includegraphics[width=3.5in]{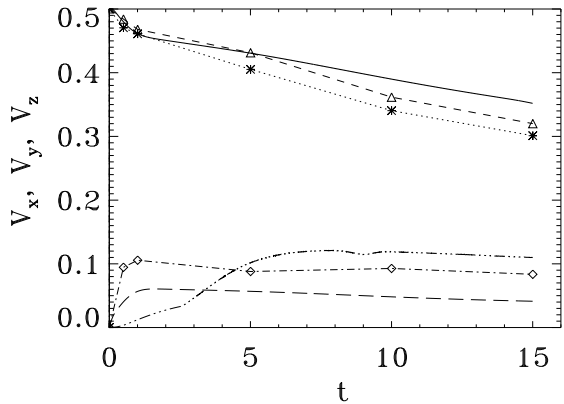}%
\caption{Evolution of maximal values of the relative 
amplitude of the generated compressible
perturbations, transverse $max(|V_x(x,z,t)|)$ and longitudinal
$max(|V_z(x,z,t)|)$ and that of Alfv\'en wave $max(|V_y(x,z,t)|)$, in time.
The solid curve presents decay of the initial Alfv\'en perturbation
due to shock dissipation in the case of no plasma density
inhomogeneity ($\lambda=0$).
The short dashed curve with triangles
depicts $max(|V_y(0,z,t)|)$ when $\lambda=0.31$, while
dotted curve with astersics represents the same physical quantity
on the edge of the similation box (away from the phase-mixing region)
i.e. $max(|V_y(15,z,t)|)$.
The dash-dotted curve with diamonds represents $max(|V_z(x,z,t)|)$,
while the dash-triple-dotted curve shows $max(|V_x(x,z,t)|)$
both for the case of $\lambda=0.31$.
The long dashed curve represents $max(|V_z(x,z,t)|)$ for the
case when $\lambda=0$.
Here $A=0.5$ and $\beta=2$.
\label{fig3}}
\end{figure}

The secular generation of fast magnetoacoustic waves by 
phase-mixed Alfv\'en waves in {\it initial} stage of the wave interaction
was observed in full-MHD numerical simulations \cite{nrm}, \cite{malara}.
The adequate modelling of the developed stage of this phenomenon
could be performed with very high numerical resolution 
that became possible only recently.
Indeed, it was found \cite{gert} that in the weakly
nonlinear regime,  phase-mixing
leads to quick saturation of the interaction.
  In the regime, 
the saturation does
not allow the fast wave amplitude to grow to more than a few tenth of a
percent of the initial  Alfv\'en wave amplitude. However, the tendency of
the saturation level to grow with the Alfv\'en wave amplitude was
noticed, which indicated that in the case of higher amplitudes,
the generation of compressible waves can be significant.
In the strongly nonlinear regime, in addition to phase mixing, coupling of MHD waves
is also affected by the steepening of
the Alfv\'en wave \cite{cke}, and by the back reaction 
of the generated fast wave on the Alfv\'en wave.
Indeed, as it has been shown by numerical simulations \cite{malara}, the  nonlinear interaction
of magnetoacoustic and Alfv\'en waves causes  energy transfer to smaller 
and smaller spatial scales. This
phenomenon may lead to the formation of shock waves and, therefore, to increased
heating of plasma by viscosity and resistivity. 
In modelling of 
non-resonant interaction of MHD waves on a plasma inhomogeneity, 
 we consider a plasma configuration similar to the
one investigated in \cite{malara} and \cite{nrm}, i.e. the plasma
has one-dimensional inhomogeneities in the equilibrium density
$\rho_0(x)$ and temperature $T_0(x)$, and is penetrated by a
straight and homogeneous magnetic field directed along the
$z$-axis. The unperturbed total pressure is taken to be constant
everywhere. The density profile is a smooth interface, 
\begin{equation}
\rho(x) =
3-2\,\mathrm{tanh} (\lambda x),
\label{profile} 
\end{equation}
where $\lambda$ is a parameter prescribing the steepness
of the profile near $x=0$. The density is normalized to $\rho_0(x=0)$. 
The sharpest gradients are located near $x=0$.  
Both the Alfv\'en speed $C_A(x)$ and the sound speed $C_s(x)$ are inhomogeneous in the
$x$-direction, however the parameter $\beta$ and the unperturbed total 
pressure are 
constant across the profile.
Dynamics of the plasma is described by single-fluid MHD equations 
in Cartesian coordinates. 
The MHD equations are solved with {\it Lare2d}  \cite{Arber}.

\begin{figure}
\includegraphics[width=3.5in]{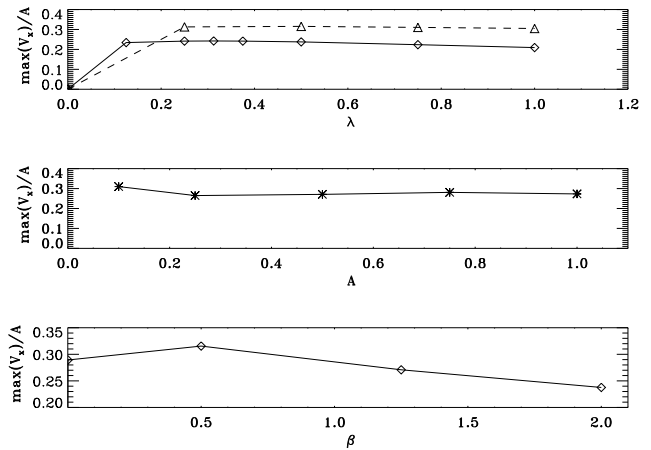}%
\caption{{\bf Top panel}: Maximal values of the relative amplitude 
of the generated compressible
perturbations for different values
of the inhomogeneity parameter
$\lambda$, with $\beta=2$ (solid curve) and $\beta=0.5$ (dashed curve).
The initial amplitude of the Alfv\'enic pulse is $A=0.5$.
{\bf Mid panel}: Maximal values of the relative amplitude of the 
generated compressible
perturbations for different values of 
initial amplitude of the Alfv\'enic pulse, with
 inhomogeneity parameter $\lambda=0.5$ and $\beta=1.25$.
{\bf Bottom panel}: Maximal values of the relative amplitude of 
the generated compressible
perturbations for different values of $\beta$, 
with initial amplitude of the 
Alfv\'enic pulse $A=0.5$ and inhomogeneity parameter $\lambda=0.5$.
\label{fig4}}
\end{figure}

An initially plane linearly polarized Alfv\'enic pulse of  Gaussian shape 
and  highly
nonlinear amplitude $V_y=0.5 C_A(0)$, 
propagates along the background magnetic field, in the $z$-direction.
{\it There are no compressible perturbations present 
in the system at the beginning.} The pulse
is subject to phase-mixing, which takes place in the vicinity of $x=0$.
Fig.~1 shows the development of the transverse gradients in the Alfv\'en pulse.
Longitudinal and transverse gradients generate compressible perturbations ($V_x$, $V_z$, $\rho$, $B_z$
and $B_x$), associated with slow and fast magnetoacoustic waves.
Note, that  initially plane Alfv\'en waves (perturbing
$V_y$ and $B_y$) in the {\it homogeneous} medium do {\it not}
generate the perturbations
of $V_x$, $B_x$ and $B_z$.
Thus, the generation of these perturbations in our simulations
are caused by the effect of plasma inhomogeneity (see Fig.~3).
It is clearly seen (Fig.~2)
that the relative perturbations of the plasma density are 
{\it enhanced by a factor of about two} 
in the phase-mixing region (near $x=0$). 
As in the weakly-nonlinear regime (\cite{gert}), initially the
compressible disturbances grow secularly, according to Eq.~(\ref{secg}), and then a 
saturated state is approached (Fig.~3). 
The saturation time is proportional to the efficiency of
the generation. However,  the maximal relative amplitude of compressible perturbations is
{\it practically independent of the problem paramteres}. The parametric study
 (Fig.~4, top and mid panel) demonstrates that the efficiency of 
 the non-resonant generation
of compressible disturbances by phase-mixed Alfv\'en waves depends weakly 
upon the initial  Alfv\'en
wave amplitude and the steepness of the inhomogeneity, 
represented by the parameter $\lambda$
(see Eq.~(\ref{profile})). The dependence on the plasma 
parameter $\beta$ is more pronounced
(see Fig.~4, bottom panel),
and the highest amplitude of compressible 
disturbances is reached for $\beta \approx 0.5$. 
{\it For all investigated parameters, the observed 
saturation level was about 30\% of the initial 
Alfv\'en wave amplitude.}

We conclude that in the presence of plasma inhomogeneity  across
the magnetic field, incompressible (Alfv\'en) and compressible
(magnetoacoustic) MHD fluctuations are effectively coupled 
 by non-resonant mechanisms. This coupling takes place
even if the waves are linearly decoupled, e.g. the perturbations
are plane in the ignorable direction perpendicular to both the
magnetic field and the inhomogeneity gradient. 
These results demonstrate that the efficiency of the nonlinear interaction of
the different type modes is strongly affected by the inhomogeneity of the
medium. In particular, in the presence of a one-dimensional inhomogeneity,
nonlinear generation of compressible disturbances by 
incompressible Alfv\'en waves
is enhanced up to a factor of two (Fig.~3). The amplitude 
of the generated compressible disturbances 
experiences saturation at the level of about 30\% of the initial amplitude
of the source Alfv\'en wave. For example, it is 10-20\% of the background
density for the Alfv\'en wave amplitude of about 0.5. 
 In particular, {\it this result is relevant to the
explanation of the absence of a significant compressible component
in the solar wind MHD turbulence}.
Indeed, the density fluctuations of less than 30\% of the background value,
prescribed by the saturation, can hardly be detected in a high-$\beta$ plasma
because of a high level of the thermal noise in the {\it in situ} data.
These results are especially relevant to the interpretation of observational data
obtained by {CLUSTER-II} mission. This issue will be discussed in more detail
elsewhere.

Finally, we would like to point out that the proposed mode coupling
mechanism could find applications in other branches such as
propagation of elastic waves in a inhomogeneous, fluid-saturated
porous medium, as mathematically, the equations describing this
phenomenon are similar to the MHD equations \cite{j}.

 Numerical calculations of this work were
done using the PPARC funded Compaq MHD Cluster in St~Andrews.

\end{document}